\documentclass[11pt]{amsart}

\usepackage[margin=1.2in]{geometry}
\usepackage{times}
\usepackage{amsmath}
\usepackage{amssymb}
\usepackage{amsfonts}
\usepackage{amsthm}
\usepackage{mathrsfs}
\usepackage[all]{xy}
\usepackage{endnotes} 
\usepackage{cite}

\usepackage[pdftex]{graphicx}
\usepackage{epstopdf}

\title{}
\author{}
\thanks{}
\date{}

\begin{document}
\bibliographystyle{nature}

\begin{titlepage}

\begin{center}
\LARGE{\textbf{Gate defined quantum confinement in suspended bilayer graphene}}\\[1.5 cm]
\large{M. T. Allen$^{1}$, J. Martin$^{1\dagger}$, and A. Yacoby$^{1*}$}\\[0.5 cm]

$^{1}$\textit{Department of Physics, Harvard University, Cambridge, MA 02138, USA }\\[0.5 cm]

$^{\dagger}$Current address: College of Engineering, Mathematics and Physical Sciences, University of Exeter, Exeter, UK, EX4 4QL %[0.5 cm]

$^{*}$email: yacoby@physics.harvard.edu
\end{center}

\end{titlepage}

\begin{abstract}

Quantum confined devices that manipulate single electrons in graphene are emerging as attractive candidates for nanoelectronics applications ~\cite{Burkard1, Stampfer1, Geim1, Dai1}. Previous experiments have employed etched graphene nanostructures, but edge and substrate disorder severely limit device functionality~\cite{DGG1, Martin2}. Here we present a technique that builds quantum confined structures in suspended bilayer graphene with tunnel barriers defined by external electric fields that break layer inversion symmetry~\cite{McCann1, Castro1}, thereby eliminating both edge and substrate disorder. We report clean quantum dot formation in two regimes: at zero magnetic field $B$ using the single particle energy gap induced by a perpendicular electric field and at $B>0$ using the quantum Hall ferromagnet $\nu=0$ gap for confinement ~\cite{Feldman1,Zhao1,Martin1}.  Coulomb blockade oscillations exhibit periodicity consistent with electrostatic simulations based on local top gate geometry, a direct demonstration of local control over the band structure of graphene.  This technology integrates single electron transport with high device quality and access to vibrational modes, enabling broad applications from electromechanical sensors to quantum bits.
\end{abstract}
\maketitle
Nanopatterned graphene devices, from field-effect transistors to quantum dots~\cite{Meric1, Stampfer1, Geim1}, have been the subject of intensive research due to their novel electronic properties and two-dimensional structure~\cite{Geim2, CastroNeto1}. For example, nanostructured carbon is a promising candidate for spin-based quantum computation~\cite{Burkard1} due to the ability to suppress hyperfine coupling to nuclear spins, a dominant source of spin decoherence~\cite{Glazman1, Petta1, Koppens1}, by using isotopically pure $^{12}$C.  Graphene is a particularly attractive host for lateral quantum dots since both valley and spin indices are available to encode information, a feature absent in GaAs~\cite{Recher1, Wu1, DasSarma1}.  Yet graphene lacks an intrinsic bandgap~\cite{CastroNeto1}, which poses a serious challenge for the creation of such devices.  Transport properties of on-substrate graphene nanostructures defined by etching~\cite{Stampfer1, Geim1} are severely limited by both edge disorder and charge inhomogeneities arising from ionized impurities in gate dielectrics~\cite{DGG1, Martin2}.  The absence of spin blockade in etched double dots is perhaps symptomatic of these obstacles~\cite{Molitor1, Liu1}.  Unzipping carbon nanotubes yields clean nanoribbon dots, but this approach cannot produce arbitrarily shaped nanostructures with tunable constrictions ~\cite{Dai1}.  Here we report fully suspended quantum dots in bilayer graphene with smooth, tunable tunnel barriers defined by local electrostatic gating.   Our technique, which artificially modifies the bandgap of bilayer graphene over nanometer scales, achieves clean electron confinement isolated from edge disorder.\\

%Nanostructured carbon is a promising candidate for spin-based quantum computation~\cite{Burkard1} due to the ability to eliminate hyperfine coupling to $^{13}C$ nuclear spins, a dominant source of spin decoherence~\cite{Glazman1, Petta1, Koppens1}

Bernal stacked bilayer graphene is naturally suited for bandgap control because of its rich system of degeneracies that couple to externally applied fields.  At $B=0$, breaking layer inversion symmetry opens an energy gap tunable up to 250 meV with an external perpendicular electric field $E$ ~\cite{McCann1, Castro1, Oostinga1, Zhang1, Xia1, Thiti1} that can be used for confinement.  In devices with low disorder and at high magnetic fields, gapped states emerge from Coulomb-driven effects that break its eightfold degeneracy (spin, valley, and orbital), resulting in quantum Hall plateaus at all integer multiples of $e^2/h$ for electron charge $e$ and Planck's constant $h$~\cite{Nomura1}.  Due to the Pauli exclusion principle, Coulomb repulsions between electrons favor spontaneous spin and/or valley polarization (or combinations of those), known as quantum Hall ferromagnetism, resulting in a gap at zero carrier density that far exceeds the Zeeman splitting energy $g\mu_B B$~\cite{Feldman1,Zhao1}.  The large exchange-enhanced energy gap of $\Delta =1.7$ meV/T measured for the $\nu=0$ state is ideally suited for quantum confinement~\cite{Martin1}.  Because valley and layer indices are identical in the lowest Landau level, one may additionally induce a tunable valley gap in the density of states by applying a perpendicular $E$ field that breaks layer inversion symmetry~\cite{Weitz1}.  This coupling of valley index to $E$ field is the key property that enables direct experimental control of the relative spin and valley gap sizes in magnetic field.    \\

We fabricate fully suspended quantum dots with 150 to 450 nm lithographic diameters as illustrated schematically in Fig.\ 1a.  Following mechanical cleavage of highly ordered pyrolytic graphite crystals, graphene is deposited on a 300 nm thermally grown $\mathrm{SiO_2}$ layer, which covers a doped silicon substrate functioning as a global back gate.  After using electron-beam lithography to define Cr/Au electrodes, an evaporated $\mathrm{SiO_2}$ spacer layer, and local top gates over selected bilayers, we immerse the samples in HF to dissolve away the dielectric on either side of the flake.  This leaves both the graphene and the top gates suspended  (Fig.\ 1a,b and Supplementary Fig. S1).  Before measurement, the devices are current annealed in vacuum to enhance quality.  The high quality of our suspended flakes is evident from the full lifting of the eightfold degeneracy in the quantum Hall regime and large resistances attained by opening the $E$ field induced gap at $B=0$ and $E=90$V/nm, a hundred times greater than reported for on-substrate bilayers at similar electric fields~\cite{Weitz1, Oostinga1}.   Measurements are conducted in a dilution refrigerator at an electron temperature of 110 mK, as determined from fits to Coulomb blockade oscillations.  \\

At $B = 0$, the electric field effect in bilayer graphene enables the production of quantum confined structures with smooth, tunable tunnel barriers defined by local gating~\cite{Recher1}, thus avoiding disorder arising from the physical edge of the flake.  Broken layer inversion symmetry opens a bandgap $\Delta \propto E = (\alpha V_t - \beta V_b)/2e\epsilon_0$, where $V_t$ and $V_b$ are top and back gate voltages with coupling factors $\alpha$ and $\beta$, respectively, and $\epsilon_0$ is vacuum permittivity. Coupling to the back gate $\beta$ is extracted from Landau fans in the quantum Hall regime and the relative gate coupling $\alpha/\beta$ can be determined from the Dirac peak slope in a $V_t$ vs. $V_b$ plot of conductance at $B=0$.  Properties of individual quantum point contacts are described in greater detail in the Supplementary Information, where pinch-off and behavior consistent with conductance quantization are observed (Supplementary Fig. S2 and S3).  Quantum dot formation at $B=0$ is illustrated schematically in Fig.\ 1c.  To create tunnel barriers beneath the top gates, we induce a bandgap by applying a field $E$ while fixing $V_t$ and $V_b$ at a ratio that maintains zero carrier density $n$, where $n=\alpha V_t + \beta V_b$.  In the non top-gated regions, there is a finite charge accumulation due to an uncompensated back gate voltage.  For gates in a quantum dot geometry, this restricts electron transport to resonant tunneling events through the constrictions.  Periodic Coulomb blockade oscillations are observed at $B=0$ which couple to both top and back gates (Fig.\ 2a).  A peak in the 2D Fourier transform corresponding to an oscillation spacing of $\sim11$mV in $V_b$ reflects this strong periodicity (Fig.\ 2b), and the appearance of higher harmonics reveals the non-sinusoidal nature of the Coulomb blockade peaks when $k_B T \ll E_C$, where $k_B$ is Boltzmann's constant, $T$ is temperature, and $E_C$ is the dot charging energy.    Coulomb diamonds shown in Fig.\ 2c have symmetric structure that suggests equal tunnel coupling to both the source and drain leads.  The dot charging energy extracted from the DC bias data is $E_C \approx 0.4$ meV.  Fig.\ 2d indicates that the periodic Coulomb blockade oscillations have comparable capacitive coupling to each pair of top gates.  \\

Coulomb blockade oscillations can also be generated at finite $B$ field using the exchange-enhanced $\nu=0$ gap.  Here the bilayer is naturally in a gapped quantum Hall state at zero density, where high resistances due to quantum Hall ferromagnetism make this system ideal for confinement.  An isolated puddle of charge is created by fixing the Fermi energy in the top-gated regions at the middle of the $\nu=0$ gap while allowing occupation of higher Landau levels elsewhere, shown schematically in Fig.\ 3a.   It should be noted that measurements in the quantum Hall regime are conducted in the valley-polarized $\nu=0$ state, far from the transition to the spin-polarized phase~\cite{Weitz1}.  Fig.\ 3b shows over forty consecutive Coulomb blockade oscillations generated at 5.2 T in a 2-gate dot with a 400 nm lithographic diameter.  The slopes of the resonances indicate symmetric coupling to the two top gates, as expected for a centrally located dot.  As top gate voltages are swept to more positive values, peak amplitude is suppressed, revealing moderate tunnel barrier tunability.    Also seen in Fig.\ 3b are interruptions in the conductance resonances (vertical and horizontal features) that couple exclusively to a single top gate; due to their sparse and aperiodic nature, we believe that they represent charging events below the gates.  Coulomb blockade oscillations are robust over a wide voltage range: Fig.\ 3c shows an additional forty peaks generated under new gate conditions.  The Coulomb diamonds exhibit symmetric tunnel coupling to source and drain leads and a dot charging energy of $E_C \approx 0.4$ meV.  The strongly periodic nature of the oscillations is evident in the Fourier transform of the data (Fig.\ 3d).  See Supplementary Fig. S4 for additional Coulomb blockade data. \\

To demonstrate geometric control over dot size, we examine the correspondence between top gate dimensions and Coulomb blockade peak spacing.  Measurements were performed on five dots with lithographic diameters ranging from 150-450 nm at magnetic fields of 0 to 7 T.  The ability to decrease peak spacing by increasing lithographic dot size is illustrated in Fig.\ 4a-c. Fig.\ 4a and 4b show Coulomb blockade peak conductance as a function of back gate voltage $V_b$ observed in device \textit{D1} at $B=5.2$ T, and \textit{D2} at $B=7$ T, respectively (see Supplementary Information for sample labeling key).  Black points represent data and the red lines are fits used to extract peak positions.  Fig.\ 4c shows relative peak position, $V(p)-V_0$, plotted as a function of peak number $p$ for the first 10 peaks of Fig.\ 4a and 4b, where $V(p)$ is the position of peak $p$ in back gate voltage, and $V_0$ is the position of the first peak.  Each data set is accompanied by a corresponding plot of $y(p)=[\frac{1}{9}\sum_{q=0}^8 V(q+1)-V(q)]p$, where $p$ and $q$ are peak index numbers, representing the average peak spacing (black lines in Fig.\ 4c).  The dot area extracted from quantized charge tunneling is given by $A = 1/(\beta\cdot\Delta V_b)$, where $\Delta V_b$ is the back gate voltage needed to increase dot occupation by one electron (Supplementary Table 1). A comparison of measured dot diameter, $d=2 \sqrt{A/ \pi}$, with effective lithographic diameter, $d_{lith}=2 \sqrt{A_{lith}/ \pi}$, indicates that $d$ generally exceeds $d_{lith}$.  This is contrary to the reduced dimensions observed in GaAs dots, where smaller dimensions are observed due to depletion~\cite{Glazman2}.\\

To obtain a better quantitative understanding of the discrepancy between lithographic and measured dot sizes, we use a commercial finite element analysis simulation tool (COMSOL) to calculate the expected dot area for each top gate geometry.  The spatial carrier density profile is modeled for a fixed top gate potential by solving the Poisson equation assuming a metallic flake in free space (Fig.\ 4d and Supplementary Fig. S5).  This assumption is justified by local compressibility measurements of the $\nu=0$ state yielding $d\mu/dn = 2 \times 10^{-17} eVm^2$ at 2 T, which translates to a screening of  99\% of the applied $V_b$ voltage by the bilayer~\cite{Martin1}.  Surprisingly, one may calculate dot size purely from gate geometry without relying on measured gap parameters.  Assuming that charge accumulation in the quantum dot occurs when the carrier density exceeds a fixed cutoff $d_0$, the dot size is defined as the area bounded by the intersection with the density distribution $f(x,y)$ with the cutoff. Assuming that the tunneling probability into the dot decays exponentially with barrier width, placement of the cutoff at the saddle points of the density profile within the constrictions enables maximal tunneling without loss of confinement.  The simulated dot area from this method, plotted in Fig. 4c (inset), is simply the area bounded by the closed contour of $f(x,y)$ at fixed density $d_0$ (Supplementary Fig. S5).  Alternatively, one may calculate dot size by imposing a cutoff equaling the measured gap width (Supplementary Table 2) and accounting for density offsets due to a displacement of the measurement voltage from the charge neutrality point (see Supplementary Information).  Remarkably, the cutoffs extracted by these two models are equivalent to within $\delta n \sim 10^{10} \mathrm{cm}^{-2}$, the carrier density fluctuations due to disorder in our suspended bilayers (Supplementary Fig. S5) ~\cite{Feldman1}.\\

Our model establishes a quantitative link between measured dot size and lithographic geometry and therefore may serve as a design tool for future bilayer nanodevices requiring submicron spatial control.  These include double dot systems which form the basis of a spin-based quantum computer~\cite{Burkard1}.  The production of suspended graphene quantum dots also enables study of coupling between quantized electronic and vibrational degrees of freedom~\cite{Leturcq1, Steele1}, with potential applications to nanoelectromechanical devices and the detection of quantized mechanical motion in a membrane~\cite{Levitov1, Schwab1, Bunch1, Chen1}. Furthermore, the combination of high sample quality with local gating enables study of edge modes that emerge at the interface of broken symmetry quantum Hall states in an environment well-isolated from edge disorder.\\

\section*{Methods}
Following mechanical exfoliation of highly oriented pyrolytic graphite crystals, graphene is deposited on a 300 nm thermally grown $\mathrm{SiO_2}$ layer, which covers a doped silicon substrate functioning as a global back gate.  Bilayer flakes are identified based on contrast to the substrate with an optical microscope and later verified through quantum Hall data.  Cr/Au (3/100 nm) electrodes are defined on selected bilayers using electron beam (ebeam) lithography, thermal evaporation, and liftoff in acetone.  A $\mathrm{SiO_2}$ spacer layer approximately 150 nm thick is deposited with ebeam evaporation after a second lithography step.  Local top gates are placed over the $\mathrm{SiO_2}$ spacers in a two step ebeam lithography process.  First small features that define the tunnel barriers and constrictions are patterned using Cr/Au of thickness 3/75 nm, and thicker support structures constructed of 3/300 nm of Cr/Au that traverse the evaporated $\mathrm{SiO_2}$ step are deposited immediately afterwards. The devices are immersed in 5:1 buffered oxide etch for 90 s and dried in a critical point dryer, which leaves both the graphene and the top gates suspended.

\section*{Acknowledgements}
The authors thank B. Feldman, O. Dial, H. Bluhm, and G. Ben-Shach for helpful discussions. This work is supported by the U.S. DOE Office of Basic Energy Sciences, Division of Materials Sciences and Engineering under award DE-SC0001819, and by the 2009 U.S. ONR Multi University Research Initiative (MURI) on Graphene Advanced Terahertz Engineering (Gate) at MIT, Harvard, and Boston University.  
Nanofabrication was performed at the Harvard Center for Nanoscale Systems (CNS), a member of the National Nanotechnology Infrastructure Network (NNIN) supported by NSF award ECS-0335765.
M.T.A. acknowledges financial support from the DOE SCGF fellowship, administered by ORISE-ORAU under contract no. DE-AC05-06OR23100.

\section*{Competing financial interests statement}
The authors declare no competing financial interests.

%\newpage
\bibliography{bibtex2} 

\newpage
\section*{Figure legends}
\textbf{Figure 1}$| \quad$\textbf{(a)} Schematic cross-section of a suspended gate-defined bilayer graphene quantum dot. Graphene is deposited on a 300 nm thermally grown $\mathrm{SiO_2}$ layer, followed by electron-beam lithography steps to define Cr/Au electrodes, an evaporated $\mathrm{SiO_2}$ spacer layer, and local top gates. Etching the dielectric on either side of the flake leaves both the flake and the top gates suspended. The electric field and carrier density profiles are controlled with back and top gate voltages $V_b$ and $V_t$, while application of a bias $V_{sd}$ across the electrodes enables transport measurements. \textbf{(b)} Scanning electron micrograph of quantum dot device similar to \textit{D4}. Bilayer graphene (not visible) is suspended between two electrodes below local top gates.  Green and blue lines indicate cross-sectional cuts in (a) and (c), respectively.  Red lines mark the estimated graphene boundaries. \textbf{(c)} Quantum dot formation at $B=0$, illustrated in a cross-sectional cut of energy vs. position. $E_C$ and $E_V$ mark the edges of the conductance and valence bands. Tunnel barriers are formed by inducing a bandgap with an external $E$ field while fixing $V_t$ and $V_b$ at a ratio that places the Fermi energy $E_F$ within the gap.  Uncompensated back gate voltage in the non top-gated regions enables charge accumulation in the dot and leads.\\

\textbf{Figure 2}$| \quad$\textbf{Coulomb blockade at $B=0$}. \textbf{(a)} Conductance map (units of $e^2/h$) of Coulomb blockade oscillations as a function of back gate voltage ($V_b$) and the voltage on top gates 1 and 2 ($V_{t12}$) at $T=110$ mK in a four gate dot (device \textit{D4}; see Fig. 1b for labeling).  The voltage on top gates 3 and 4 is fixed at $V_{t34}=9.27$ V. \textbf{(b)} 2D fast Fourier transform of (a) reveals the periodic structure.  A peak corresponding to an oscillation spacing of $\sim11$mV in $V_b$ reflects strong periodicity, while the appearance of higher harmonics reveals the non-sinusoidal nature of the Coulomb blockade peaks when $k_B T \ll E_C$. \textbf{(c)} Coulomb diamonds are shown in a plot of $\Delta G / \Delta V_{t12}$as a function of $V_{t12}$ and $V_{DC}$, where $G$ is conductance in units of $e^2/h$ and $V_{DC}$ is the DC bias across the electrodes.  The voltages $V_b=-10.7$V and $V_{t34}=9.27$ V are held constant.  Symmetric Coulomb diamonds suggests equal tunnel coupling to source and drain leads.  The dot charging energy is $E_C \approx 0.4$ meV.  \textbf{(d)} Conductance map (units of $e^2/h$) of Coulomb blockade oscillations as a function of $V_{t12}$ and $V_{t34}$ at fixed back gate voltage $V_b=-10.7$V.\\

\textbf{Figure 3}$| \quad$\textbf{Coulomb blockade at $B=5.2$ T}. \textbf{(a)} Quantum dot formation at $B>0$, illustrated in a cross-sectional cut of energy vs. position, following the red line in Fig. 1b. Tunnel barriers are formed using the exchange-enhanced $\nu=0$ gap, where high resistances due to quantum Hall ferromagnetism are ideal for confinement.  An isolated puddle of charge is created by fixing the Fermi energy in the top-gated regions at the middle of the $\nu=0$ gap while allowing occupation of higher Landau levels elsewhere. \textit{Inset}: Schematic illustration of the top gate geometry for device \textit{D1}.   \textbf{(b)} Conductance map (units of $e^2/h$) of Coulomb blockade oscillations as a function of $V_{t1}$ and $V_{t2}$ in a two top-gate dot (device \textit{D1}), at fixed back gate voltage $V_b=-15.4$V and $T=110$ mK. The slopes of the resonances indicate symmetric coupling to the two top gates, as expected for a centrally located dot.  As top gate voltages are swept to more positive values, peak amplitude is suppressed, revealing moderate tunnel barrier tunability.  \textbf{(c)} Coulomb diamonds are shown in a plot of conductance (units of $e^2/h$) as a function of $V_{t1}$ and DC bias $V_{DC}$, where $V_{t2}=11$ V and $V_b=-14.4$ V are fixed.  Symmetric Coulomb diamonds suggests equal tunnel coupling to source and drain leads.  The dot charging energy is $E_C \approx 0.4$ meV. \textbf{(d)} 2D fast Fourier transform of the boxed region in (b), revealing the strongly periodic nature of the oscillations and higher harmonics.\\

\textbf{Figure 4}$| \quad$\textbf{Geometric control over Coulomb blockade period}. \textbf{(a)} Coulomb blockade peak conductance as a function of back gate voltage $V_b$ observed in device \textit{D1} at $B=5.2$ T.  Black points represent data and the red line indicates a functional fit used to extract peak positions.  The top gate voltages are fixed at $V_{t1}=11.402$ V and $V_{t2}=12$ V.\textbf{(b)} Coulomb blockade in device \textit{D2} at $B=7$ T (See Fig. S1c in the Supplementary Information for labeling).  The top gate voltages are fixed at $V_{t1}=V_{t3}=13$ V and $V_{t2}=V_{t4}=12$ V. \textbf{(c)}  Relative peak position, $V(p)-V_0$, plotted as a function of peak number $p$ for the first 10 peaks of (a) and (b).  $V(p)$ is the position of peak $p$ in back gate voltage $V_b$, and $V_0$ is the position of the first peak ($V_0=-15.4907$ V and $-17.9875$ V for plots (a) and (b), respectively).  Each black line is a plot of $y(p)=[\frac{1}{9}\sum_{q=0}^8 V(q+1)-V(q)]p$, where $p$ and $q$ are peak index numbers, representing the average peak spacing for the particular data set. \textit{Inset}: Simulated dot size versus measured size. Error bars represent the range of diameters expected for measured Coulomb blockade peak spacings within one standard deviation of the mean. \textbf{(d)} COMSOL simulation of density profile (in arbitrary units) for the lithographic gate pattern of device \textit{D4} for top gate voltage $V_{t1}=V_{t2}=12$ V. Electron transport is restricted to resonant tunneling events through the constrictions, indicated by the arrows.  

\newpage
%Figure 1
\begin{figure}[!h]
\includegraphics[width=160mm]{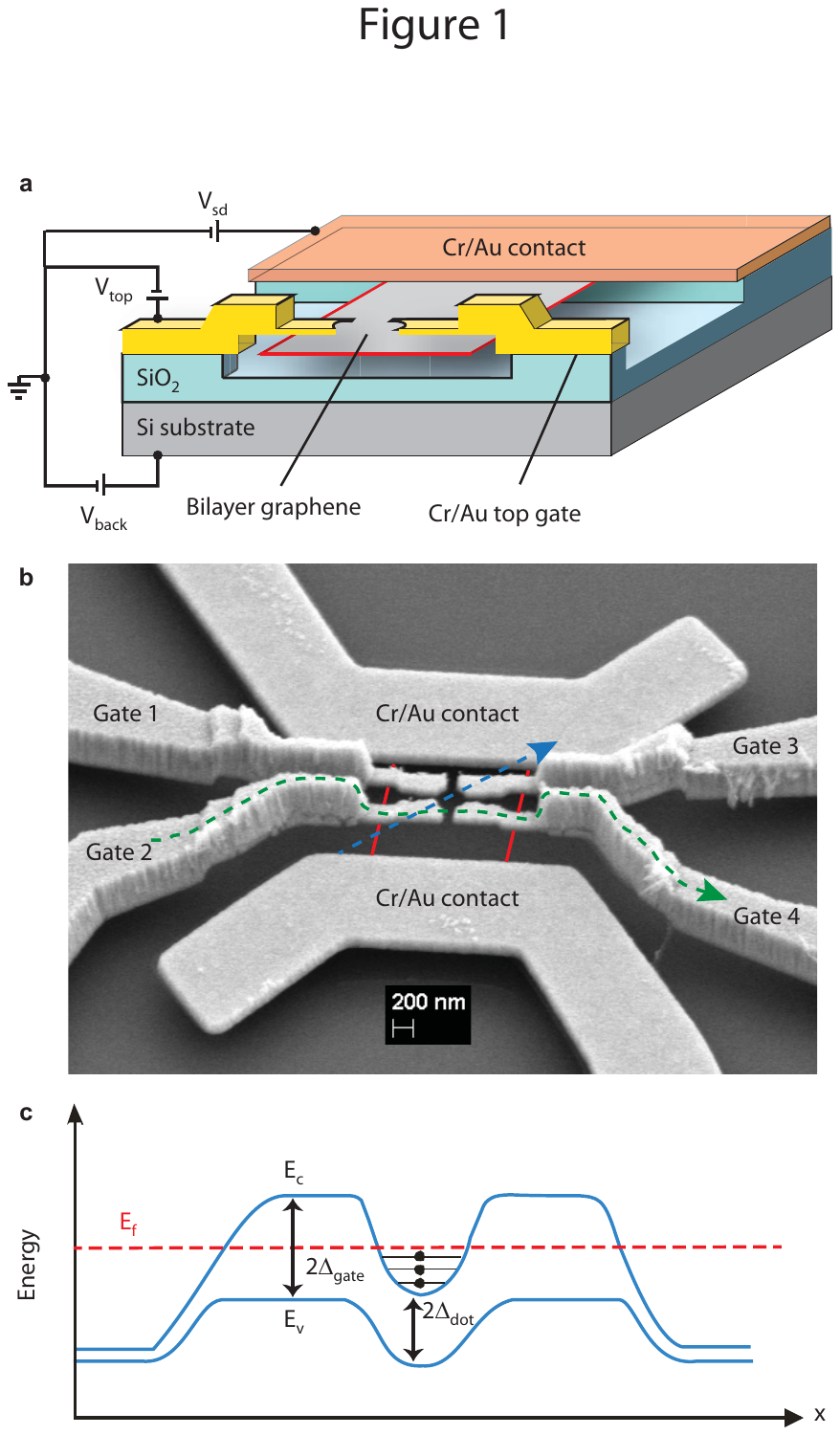}
%\caption{  } \label{fig1}}
\end{figure}

\newpage
%Figure 2
\begin{figure}[!h]
\includegraphics[width=160mm]{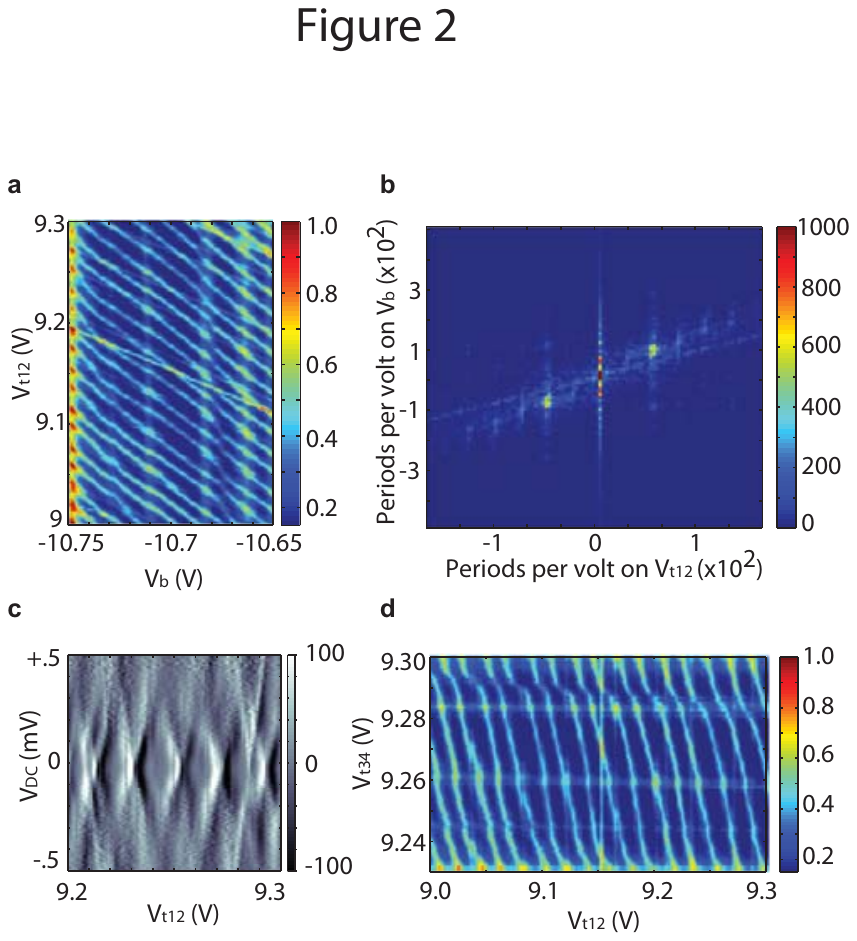}
%\caption{  } \label{fig1}}
\end{figure}

\newpage
%Figure 3
\begin{figure}[!h]
\includegraphics[width=160mm]{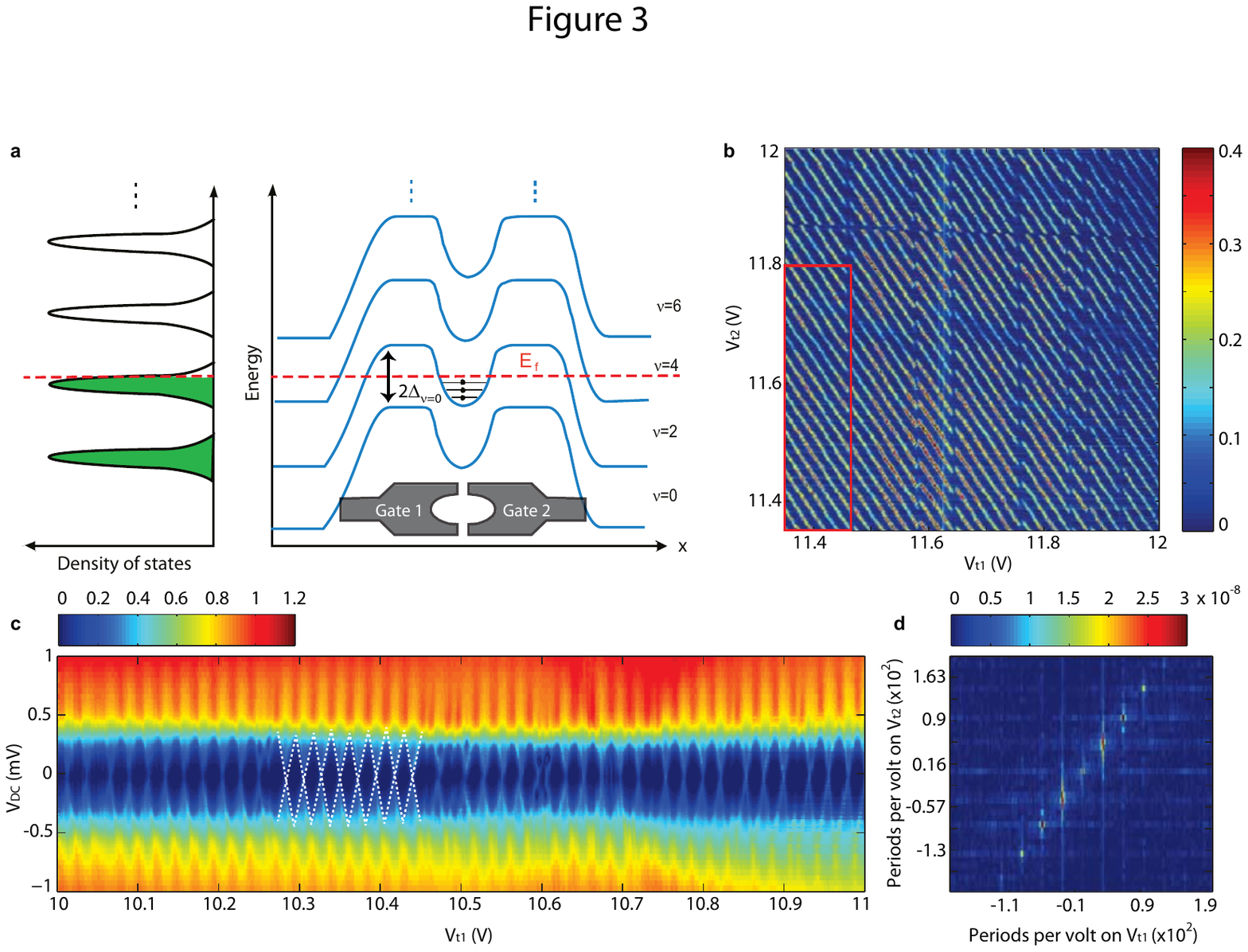}
%\caption{  } \label{fig1}}
\end{figure}

\newpage
%Figure 4
\begin{figure}[!h]
\includegraphics[width=160mm]{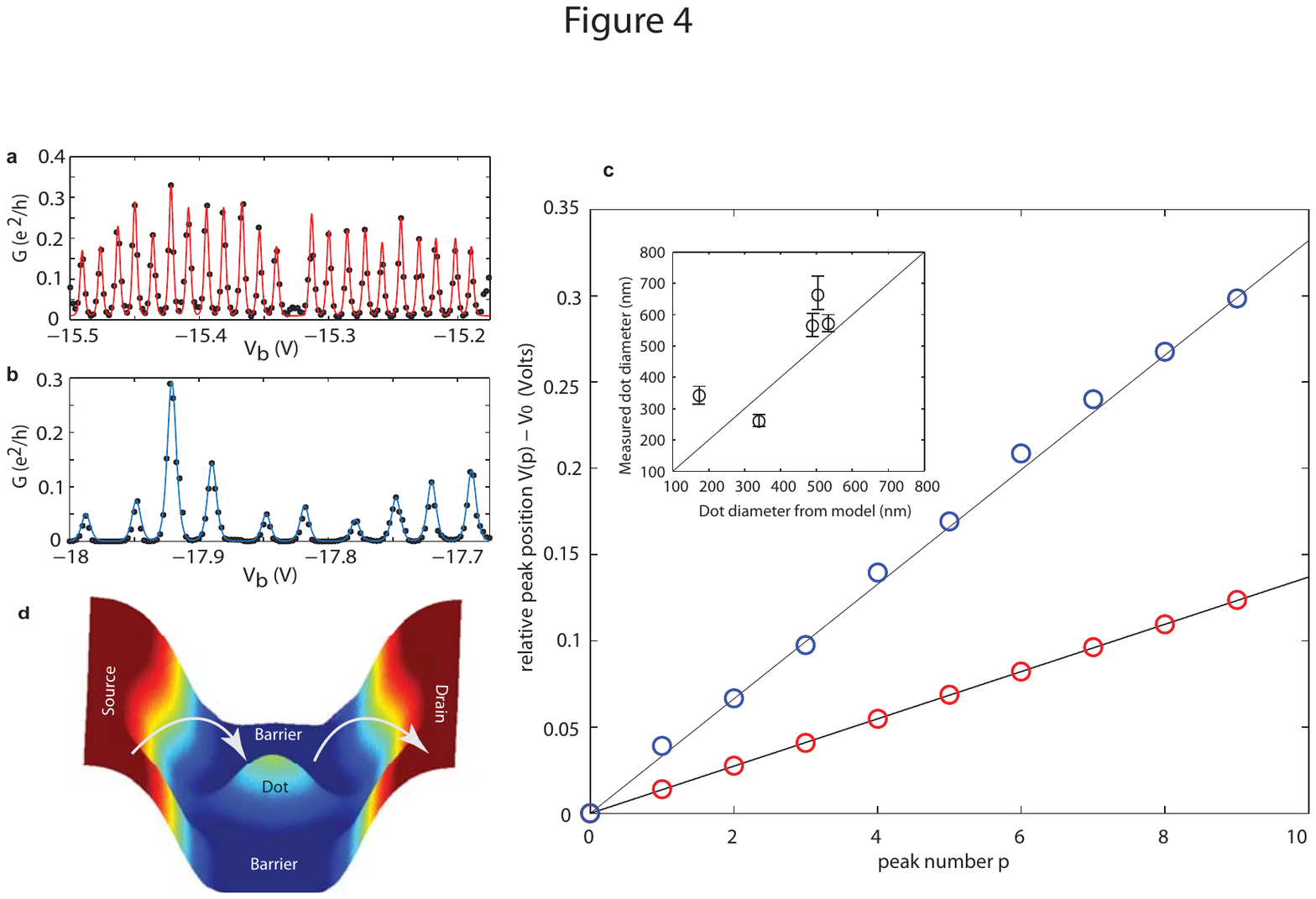}
%\caption{  } \label{fig1}}
\end{figure}

\end{document}

\bibliographystyle{//Yacobyserver/ygroup/Monica/nature}

\begin{center}
\Large{\textbf{Gate defined quantum confinement in suspended bilayer graphene}}\\[0.5 cm]
\large{M. T. Allen, J. Martin, and A. Yacoby}\\[0.5 cm]
\end{center}

\maketitle

\tableofcontents

\section{Gate defined quantum dots: device overview}
%include more SEM images

We analyze five fully suspended quantum dots with lithographic diameters, $d_{lith}=2 \sqrt{A_{lith}/ \pi}$, of 150 to 450 nm.  Listed here are details on individual sample geometries and measurement conditions.  Device \textit{D1} is a circular two gate dot with $d_{lith}=$ 400 nm measured at $B=5.2$ T with the valley polarized $\nu=0$ gap used for confinement.  Device \textit{D2} is a circular dot with $d_{lith}=$150 nm consisting of three main gates with a plunger gate suspended above (see Supplementary Figure S1c for labeling).   Measurements were conducted at $B=0$ and $7$ T with the $E$ field induced and $\nu=0$ gaps used for confinement, respectively. Device \textit{D3} is a circular four gate dot with $d_{lith}=$ 300 nm measured at $B=7$ T.  Device \textit{D4} is a circular four gate dot with $d_{lith}=$ 400 nm measured at $B=0$ and $3$ T.  Device \textit{D5} is a two gate elliptical dot with cross sectional lengths of 200 and 250 nm measured at $B=5$ T.  Additional scanning electron microscope (SEM) images of locally gated quantum dot devices are provided in Fig.\ S1.
All measurements are conducted using standard Lockin techniques in a Leiden Cryogenics Model Minikelvin 126-TOF dilution refrigerator.  An electron temperature of $T=110$ mK is extracted from Coulomb blockade fits.\\

\section{Quantum point contacts defined by local gating}
By opening a bandgap beneath local top gates in a quantum point contact (QPC) geometry (Fig.\ S2a), electron transport is restricted to conductance through the constriction.  Fig.\ S2b is a SEM image of a fully suspended QPC with 200 nm between the split gates.  Beneath the split side gates, one may induce a bandgap by applying a perpendicular $E$ field, meanwhile fixing the top and back gate voltages, $V_{t1}$ and $V_b$, at a ratio that places the Fermi energy near the center of the gap.  Application of voltage $V_{t2}$ to the central gate independently tunes the carrier density within the channel, enabling one to sweep the Fermi wavelength for a fixed constriction width.  Full pinching off of the constriction is demonstrated in Fig.\ S2c.

Figures S2d and S3 contain plots of conductance in units of $e^2/h$ versus central gate voltage (labeled $V_{t2}$) in a device similar to that pictured in Fig.\ S2b. Measurements in a parallel field of $B = 5$ T show conductance steps at every integer ($e^2/h$ spacing).  The plot of differential conductance as a function of central gate voltage and magnetic field $B$ shows the emergence of odd-numbered plateaus as spin degeneracy is broken near 4-5 T (Fig.\ S2d).  A differential conductance map as a function of DC bias and central gate voltage exhibits behavior consistent with conductance quantization, though the features are obscured by Fabry-Perot like interference patterns (Fig.\ S3).

\section{Measured dot size extracted from Coulomb blockade fits}

The dot area extracted from quantized charge tunneling is given by $A = 1/(\beta\cdot\Delta V_b)$, where $\Delta V_b$ is the back gate voltage needed to increase dot occupation by one electron and $\beta$ is the carrier density induced by 1 V on the back gate. The global density to voltage conversion is obtained by fitting Landau level filling factors $\nu=nh/eB$ from bulk quantum Hall data. The peak spacing $\Delta V_b \equiv V_{b,res}^{i+1}-V_{b,res}^{i}$ of a given data set is extracted by fitting each Coulomb blockade oscillation to the conductance expression:
\begin{equation}
G(V_b)=A\cdot\cosh\left( \frac{ ea|V_b-V_{b,res}|}{ 2.5k_B T}  \right)^{-2}
\end{equation}
where $e$ is the electron charge, $k_B$ is Boltzmann's constant, $T$ is temperature, $A$ is peak amplitude, $V_{b,res}$, is the back gate voltage at resonance, and $a=C_g/(C_l+C_r+C_g)$ is determined from the slopes of the Coulomb diamonds ($C_g$ is capacitance to the back gate and $C_l$ and $C_r$ are capacitances across the left and right tunnel barriers, respectively)~\cite{Sohn1}.  This expression is valid in the regime $\Delta E \ll k_BT \ll e^2/C$, where $\Delta E$ is the single particle level spacing.  This functional fit to the data is shown explicitly in Fig.\ 4a,b.  Coulomb blockade data from an additional device is provided in Supplementary Figure S4. \\

\begin{table}[h]
\begin{tabular}{|c|c|c|c|c|}
\hline
Device  & $\beta$, $m^{-2}V^{-1}$ &  $\langle \Delta V_b \rangle$, mV & Standard deviation of $\Delta V_b$ & Dot diameter, $d=2 \sqrt{A/ \pi}$\\ %$d= 2/ \sqrt{\pi \beta\cdot  \Delta V_b }  $
\hline \hline
D1 & $2.85 \times 10^{14}$ & 14.0 & 1.8 & $565 (+40/-34)$ nm\\ %MA111710a P13
D2 & $3.19 \times 10^{14}$& 34.1 & 5.3 & $342 (+30/-24)$ nm \\ %MA062810a P13
D3& $3.05 \times 10^{14}$ & 68.8  & 4.2 & $260 (\pm 8)$ nm \\ %MA062810a P11
D4 & $2.71 \times 10^{14}$ & 10.7 & 1.7 & $663 (+60/-47)$ nm\\ %MA111710a P1 QD
D5 & $2.85 \times 10^{14}$ & 13.7 & 1.3 & $571 (+29/-26)$ nm\\ %MA111710a P14
\hline 
\end{tabular}
\caption{ \footnotesize{List of measured dot sizes obtained from Coulomb blockade fits for five different samples. $\beta$ is the carrier density induced by 1 V on the back gate and $\langle \Delta V_b \rangle$ is the average peak spacing. Dot diameter is expressed as $d=2 \sqrt{A/ \pi}$ for area $A = 1/(\beta\cdot\Delta V_b)$. Error bars in the last column represent the range of diameters expected for measured Coulomb blockade peak spacings within one standard deviation of the mean. Results are plotted in Fig.\ 4a in the paper. } }
\end{table}

\section{Model 1: Cutoff defined by constriction saddle points}
We first discuss simulation of the spatial density profile for each quantum dot, as shown in Fig.\ 4d, which is relevant to the final two sections of the Supplementary Information.  COMSOL Multiphysics, a commercial finite element analysis simulation tool, is used to model the spatial carrier density profile for a fixed top gate potential by solving the Poisson equation, $\nabla^2 V = -\rho / \epsilon_0$.  First the lithographic top gate pattern, designed using TurboCAD, is imported into COMSOL and placed in a parallel plane defined 150 nm above the graphene flake.  The flake is assumed to be a two dimensional metallic plate, which is justified by local compressibility measurements of the $\nu=0$ state yielding $d\mu/dn = 2 \times 10^{-17} eVm^2$ at 2 T, which translates to a screening of  99\% of the applied $V_b$ voltage by the bilayer (for $dn= 1.5\times10^{9}$ $cm^{-2}$) ~\cite{Martin1}.  A fixed potential matching the experimental value at which Coulomb blockade oscillations appear is assigned to the top gates, while the flake is grounded.  The perpendicular electric field component, $E_z(x,y)$, is solved in a plane 5 nm above the flake.  The approximate carrier density profile is given by the function $f(x,y)=e\cdot E_z(x,y) / \epsilon_0$ up to a constant offset.  This is because the graphene screens all in-plane electric fields.  A contour plot representing the simulated spatial density profile in device \textit{D4} at a top gate voltage of 12 V is presented in Supplementary Figure S5. 

In the first modeling approach, whose results are presented in Fig.\ 4c, one may calculate dot size purely from gate geometry without relying on measured gap parameters.  Assuming that charge accumulation in the quantum dot occurs when the carrier density exceeds a fixed cutoff $d_0$, the dot size is defined as the area bounded by the intersection with the density distribution $f(x,y)$ with the cutoff.  Assuming that the tunneling probability into the dot decays exponentially with barrier width, placement of the cutoff at the saddle points of the density profile within the constrictions enables maximal tunneling without loss of confinement. Thus, the simulated dot area computed by this method is simply the area bounded by the closed contour of $f(x,y)$ at fixed density $d_0$.

The model is used to estimate the extent to which dot size should change in response to a changing top gate voltage.  The dot diameter, $d=2 \sqrt{A/ \pi}$, for sample \textit{D2} is computed to be 170 nm at top gate voltage $V_t=9$ V and 173.5 nm at $V_t=13$V.  This 3.5 nm increase in diameter, a 2 percent change, over a 4 V range is substantially smaller than the experimental error bars due to fluctuations in peak spacing (Fig.\ 4).  This prediction is consistent with the overall experimental observation of an approximately constant dot size over the measurement ranges presented in this paper.

\section{Model 2: Cutoff determined from measured $\nu=0$ gap and offset from charge neutrality point}

In the second modeling approach, we determine quantum dot dimensions using both the simulated density distribution and experimental gap measurements. First the spatial density profile induced by the top gates, $f(x,y)$, is modeled using COMSOL following the procedure described in the preceding section of the Supplementary Information.  To determine the quantum dot size from the density profile, we impose a cutoff $d_1$ determined by the width of the gap ($\nu=0$ or $E$ field induced) above which charge accumulation begins.  Explicitly, $d_1=(V_+ -V_-) \beta$, where  $V_+$ and $V_-$ are the positive and negative back gate voltages at which a plot of conductance versus $V_b$ intersects $1 e^2/h$.  Additionally we account for overall offsets in density due to a displacement of the measurement voltage from the charge neutrality point.  This offset is given by 
$d_{offset}=(V_{meas}-V_{CNP})\beta$, where $V_{meas}$ is the measured back gate voltage at which Coulomb blockade oscillations appear, and $V_{CNP}$ is the measured back gate voltage at which the charge neutrality point appears at $B=0$ when the top gates are fixed at the potential defined in the density profile simulation.  Thus, the spatial density profile with proper offsets included is
$g(x,y)=f(x,y) - d_{sat} +d_{offset}$, where $d_{sat}$ is the saturating value of $f(x,y)$ deep within the tunnel barriers defined by the top gates.  Physically $d_{sat}$ is the offset in carrier density induced by the back gate that places the Fermi energy at the center of the bandgap.  Similar to Model 1, the simulated dot area computed by this method is the area bounded by the contour lines of $g(x,y)$ at fixed density $d_1$.  Results of Models 1 and 2 are presented in Table 2 and Supplementary Figure S5.

\begin{table}[h]
\begin{tabular}{|c|c|c|c|c|}
\hline
Device  & $d_0$, $m^{-2}$&  $d_1$, $m^{-2}$ & Dot diameter (Model 1) &  Dot diameter (Model 2) \\ 
\hline \hline
D1 & $7.87 \times 10^{14}$ & $6.81 \times 10^{14}$ & $489$ nm & $564$ nm\\ %MA111710a P13
D2 & $1.85 \times 10^{14}$& $1.45 \times 10^{14}$ & $170$ nm & $204$ nm \\ %MA062810a P13
D3& $7.59 \times 10^{14}$ & $8.34 \times 10^{14}$  & $340$ nm & $300$ nm \\ %MA062810a P11
D4 & $4.29 \times 10^{14}$ & $3.90 \times 10^{14}$ & $504$ nm & $537$ nm\\ %MA111710a P1 QD
D5 & $6.96 \times 10^{14}$ & $6.59 \times 10^{14}$ & $533$ nm & $562$ nm\\ %MA111710a P14
\hline 
\end{tabular}
\caption{ \footnotesize{List of simulated dot sizes obtained from Models 1 and 2 for five different samples. Carrier densities $d_0$ and $d_1$ represent the cutoff values above which charge accumulation begins in Models 1 and 2, respectively.  Dot diameter is expressed as $d=2 \sqrt{A/ \pi}$ for area $A$.  Diameters from Model 2 were computed at the maximum value of offset  $d_{offset}=(V_{meas}-V_{CNP})\beta$.  Results are plotted in Supplementary Figure S5. } }
\end{table}

\bibliography{bibtex_supp2}

\section*{Supplementary figure legends}
\textbf{Figure S1} $| \quad$ \textbf{Scanning electron micrographs of gate defined quantum dots in graphene} \textbf{(a)} Tilted SEM image of a four gate quantum dot device. \textbf{(b)} SEM image of a three gate quantum dot taken with the Inlens detector. An additional plunger gate used to control the density in the dot is suspended above the lower gates.  The suspended graphene bilayer is faintly visible below the gates.  \textbf{(c)} Colored SEM image of a device similar to that pictured in (b) taken with the SE2 detector.  Red lines mark the estimated graphene boundaries.  The gate geometry is nearly identical to that of device \textit{D2}, and the top gate labeling that accompanies the data in Fig. 4b of the paper is provided here.\\

\textbf{Figure S2}$| \quad$  \textbf{Graphene quantum point contacts defined by local gating} \textbf{(a)} Tilted false-color SEM image of three suspended quantum point contact (QPC) devices in series. \textbf{(b)} SEM image of fully suspended QPC with 200 nm between split gates (labeled with blue arrows).  The central gate (labeled with a pink arrow) tunes the carrier density in the channel. \textbf{(c)} Pinching off of the constriction at $B=0$ is illustrated.  This is a plot of conductance (in units of $e^2/h$) versus central gate voltage, $V_{t2}$.  The split gates are fixed at $V_{t1}=-9.75$ V and the back gate is at $V_b=11.4$ V. \textbf{(d)} Plot of conductance in units of $e^2/h$ versus central gate voltage, $V_{t2}$, in a device similar to that pictured in part b.  The split gates are fixed at $V_{t1}=-10$ V and the back gate is at $V_b=11.9$ V.  At $B=0$ conductance steps are visible at values of 4, 6, and 8 $e^2/h$ (bottom panel), while steps energe at integer multiples of $e^2/h$ in the presence of an in-plane magnetic field of $B=5$ T (top panel). This behavior is suggestive of a broken valley degeneracy at $B=0$ and the gradual breaking spin degeneracy with increasing magnetic field.  The central panel, a plot of $\Delta G / \Delta V_{t2}$, shows the gradual emergence of the integer steps over a 5 T field range.\\

\textbf{Figure S3}$| \quad$ \textbf{Conductance through a quantum point contact at nonzero source-drain bias} \textbf{(a)} Schematic of DC bias behavior for a QPC.  The black lines in the bottom panel represent transitions between conductance plateaus as a function of $V_{gate}$ and $V_{DC}$.  The energy diagrams in the upper panels show placement of the source and drain chemical potentials ($\mu_s$, $\mu_d$) relative to the one-dimensional subbands at the locations marked with blue circles. \textbf{(b)} Plot of conductance in units of $e^2/h$ versus central gate voltage (labeled $V_{t2}$) in a device similar to that pictured in Fig.\ S2b.  Steps emerge at integer multiples of $e^2/h$ in the presence of an in-plane magnetic field of $B=5$ T (top panel). A map of $\Delta G / \Delta V_{t2}$ as a function of DC bias and central gate voltage exhibits behavior consistant with conductance quantization, though the features are obscured by Fabry-Perot like interference patterns.\\

\textbf{Figure S4}$| \quad$ \textbf{Quantum confinement in device \textit{D3} } \textbf{(a)} Coulomb blockade peak conductance vs. back gate voltage $V_b$ at $B=7$ T.  Top gate voltages are $V_{t1}=V_{t3}=11$ V and $V_{t2}=10.5$ V. \textbf{(b)} Peak spacings for the data in part a. \textbf{(c)} Coulomb diamonds in a plot of $\Delta G / \Delta V_{b}$, where $G$ is conductance in units of $e^2/h$ and $V_{DC}$ is the DC bias across the contacts. \textbf{(d-f)} Conductance($e^2/h$) of Coulomb blockade peaks vs. $V_{t1}$, $V_{t2}$, and $V_{t3}$, respectively. Similar coupling to each top gate suggests a centrally located dot. \textbf{(g)} False-color scanning electron micrograph of a dot similar to \textit{D3}. Voltages $V_{t1}$, $V_{t2}$, and $V_{t3}$ are applied to pins 3, 15, and 9$\&$10, respectively.  The gates connected to pins 9 and 10 are shorted together. \\

\textbf{Figure S5}$| \quad$  \textbf{Quantum dot size determined by simulations} \textbf{(a)} Contour plot representing COMSOL simulation of spatial density profile in device \textit{D4} at a top gate voltage of 12 V.  The red line (indicated by the black arrow) is the contour line at the saddle points of the density profile, and the area bounded by the closed portion of this curve represents the quantum dot size calculated by Model 1. \textit{Inset}: Cross sectional cut of density profle along $y=0$.  The points of intersection with the cutoff (red circles) coincide with the red contour line that determines the dot area. \textbf{(b)} Simulated dot size versus measured size.  Circles and squares represent areas calculated using Models 1 and 2, respectively. Error bars represent the range of diameters expected for measured Coulomb blockade peak spacings within one standard deviation of the mean. \textbf{(c)} Comparison between cutoffs in the two modeling approaches. Carrier densities $d_0$ and $d_1$ represent the cutoff values above which charge accumulation begins in Models 1 and 2, respectively.  The black dashed lines are plots of $y=x\pm\delta n$, where $\delta n \sim 10^{14}$ $m^{-2}$ is the density variation due to disorder in our suspended flakes \cite{Feldman1}.\\

\newpage
%Figure 1
\begin{figure}[!h]
\includegraphics[width=170mm]{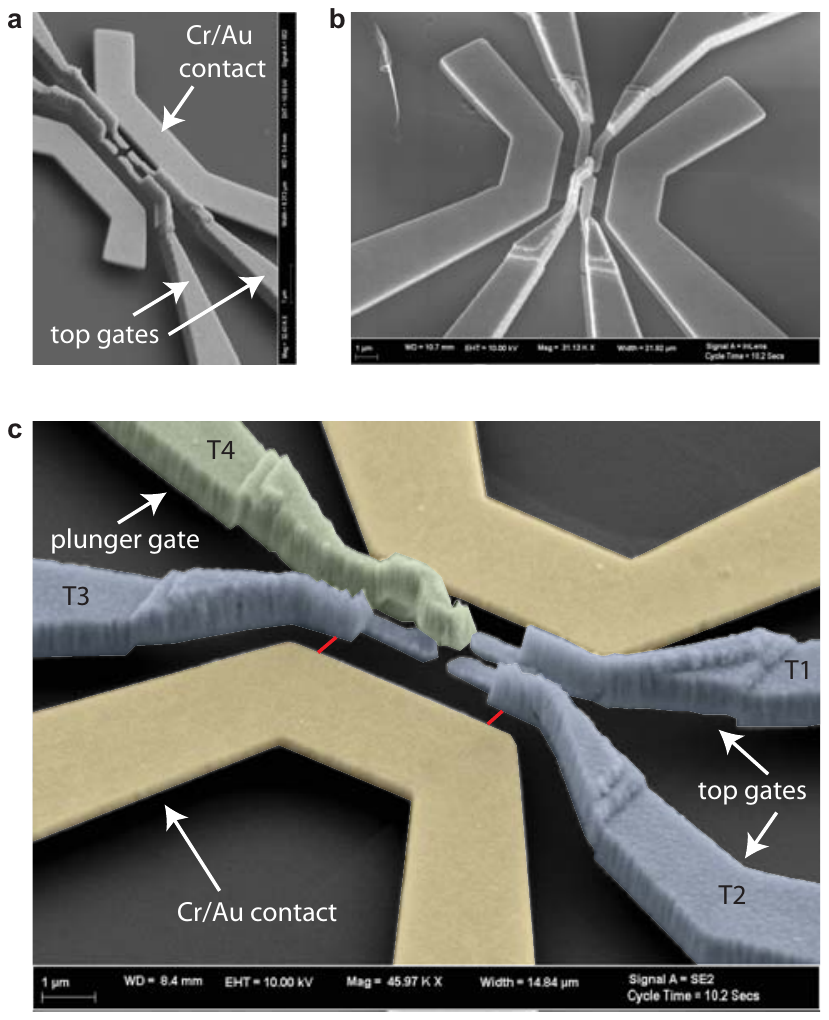}
\Large{\textbf{Figure S1}}
%\caption{} \label{fig1}}
\end{figure}

\newpage
%Figure 2
\begin{figure}[!h]
\includegraphics[width=170mm]{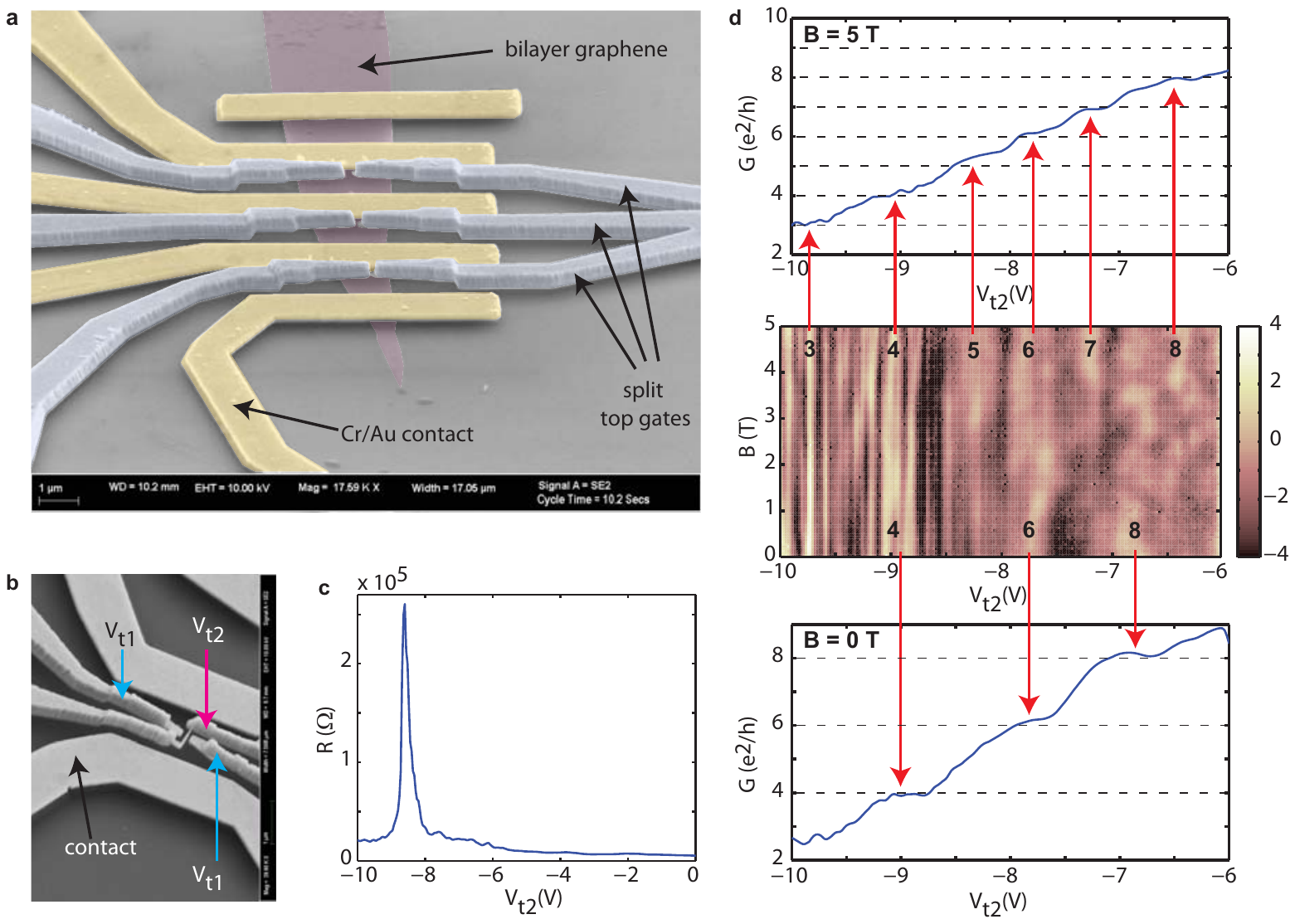}
\Large\textbf{Figure S2}
%\caption{  } \label{fig1}}
\end{figure}

\newpage
%Figure 3
\begin{figure}[!h]
\includegraphics[width=170mm]{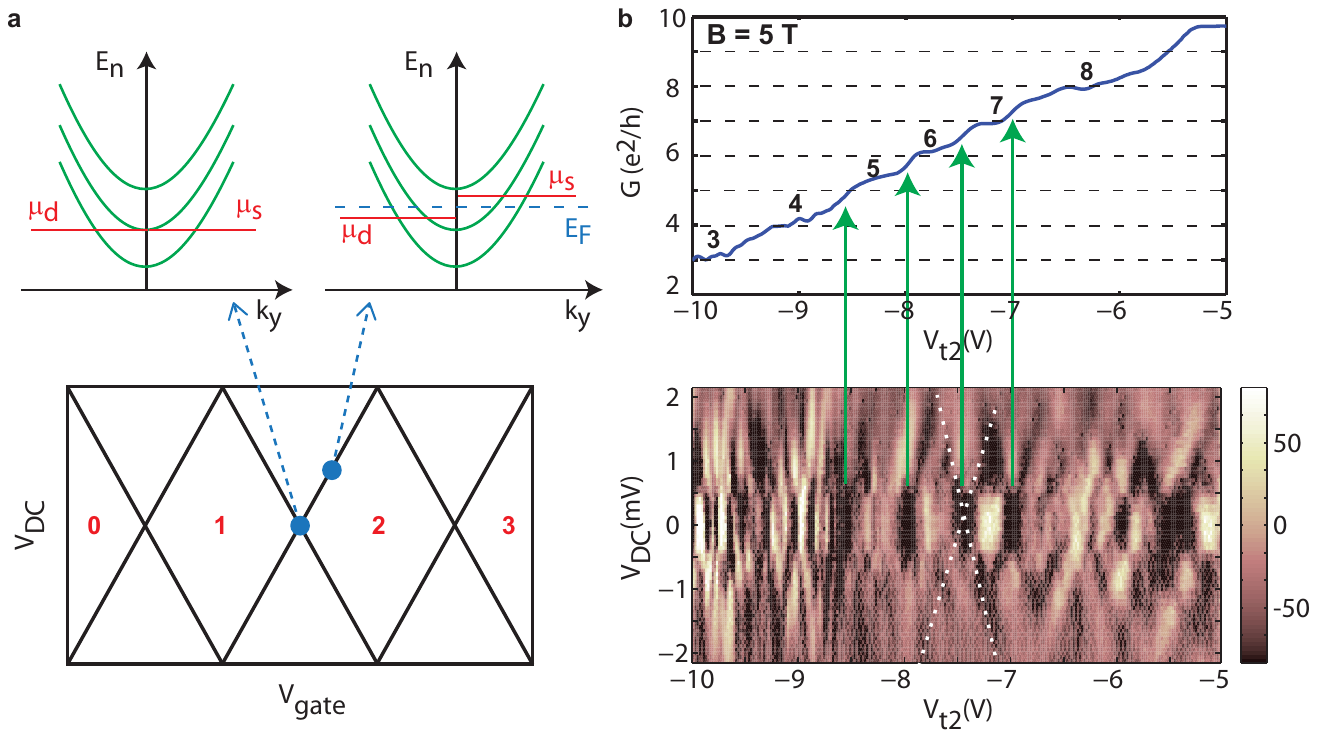}
\Large\textbf{Figure S3}
%\caption{  } \label{fig1}}
\end{figure}

\newpage
%Figure 4
\begin{figure}[!h]
\includegraphics[width=170mm]{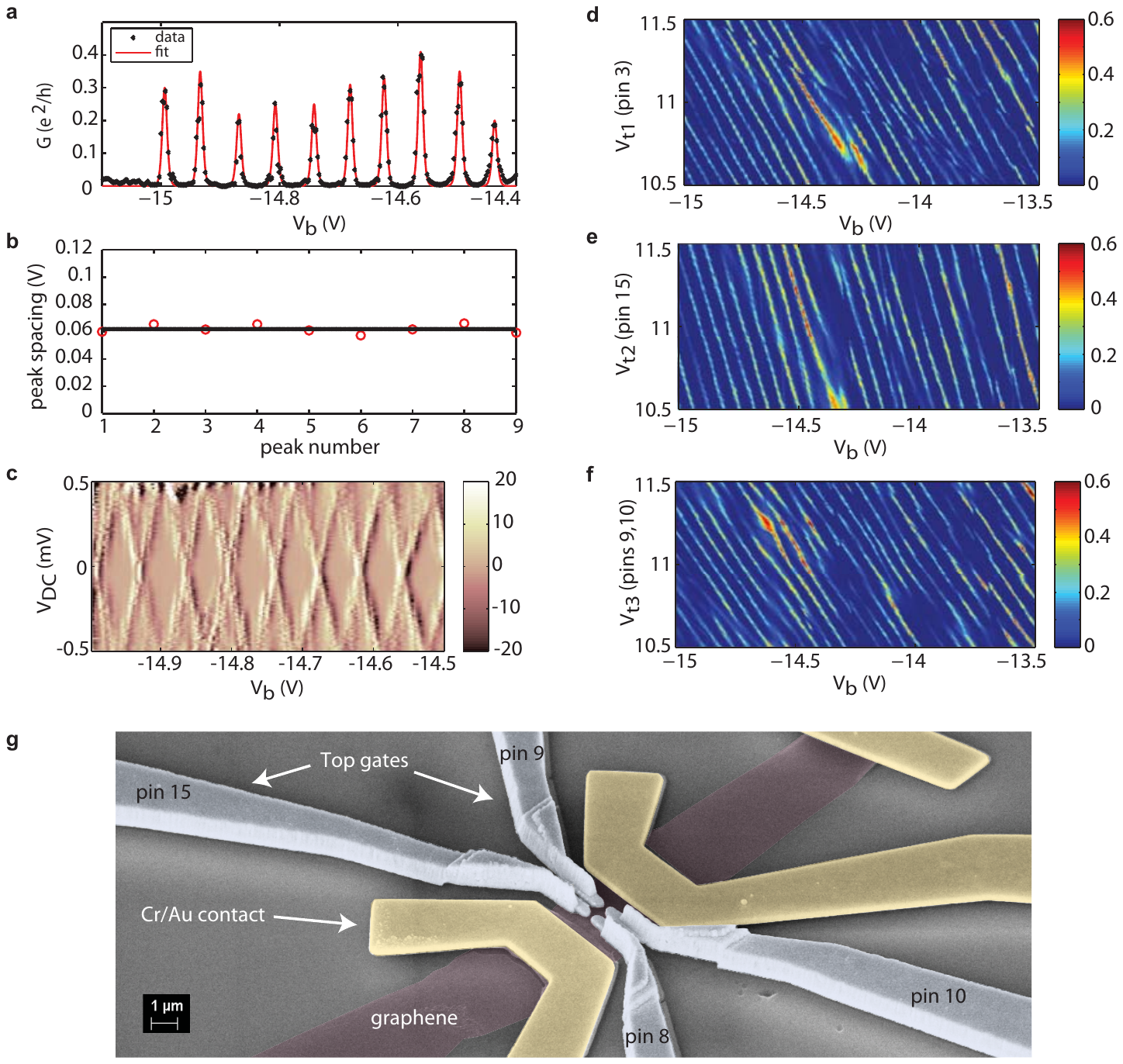}
\Large\textbf{Figure S4}
%\caption{  } \label{fig1}}
\end{figure}

\newpage
%Figure 5
\begin{figure}[!h]
\includegraphics[width=170mm]{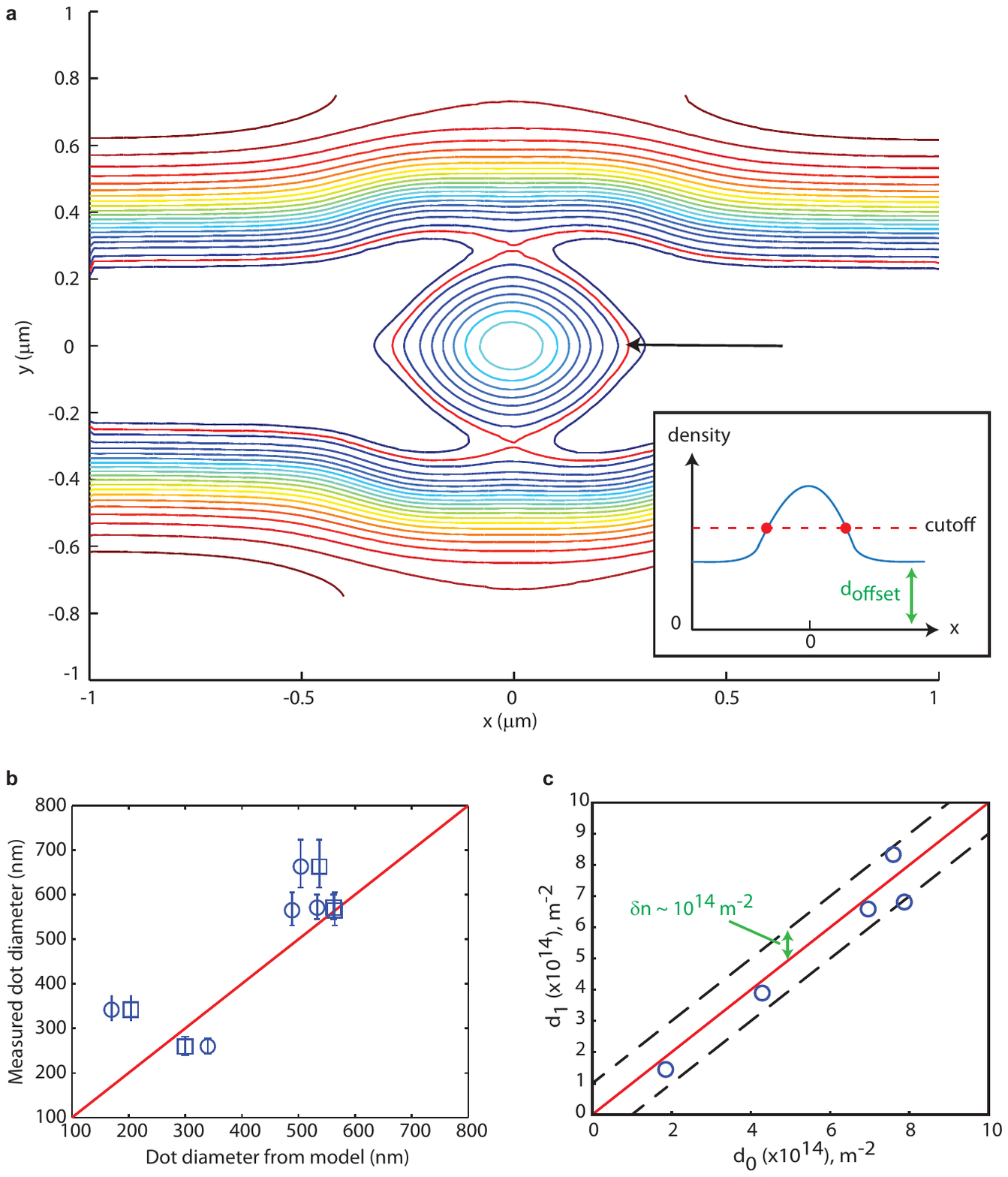}
\Large\textbf{Figure S5}
%\caption{  } \label{fig1}}
\end{figure}